# Impact of Network Densification on the Performance of a Non-Public URLLC Factory Network

Kimmo Hiltunen[1], Yanpeng Yang[2], Fedor Chernogorov[1]
[1]Ericsson Research, Jorvas, Finland
[2]Ericsson Research, Stockholm, Sweden
{kimmo.hiltunen, yanpeng.yang, fedor.chernogorov}@ericsson.com

*Abstract*—Densification of the network deployment, for example by adding new sectors or sites within an existing mobile communication network, has traditionally been an efficient way to improve the system coverage and capacity. That will be the case even for the ultra-reliable low-latency communication (URLLC) services, but the overall situation and the feasibility of the different deployment options will depend on the characteristics of the URLLC service in question. URLLC services with relaxed latency requirements, allowing multiple transmission attempts, can tolerate a decent level of inter-cell interference while still being able to guarantee the desired quality-of-service for all users, which makes it possible to improve the URLLC service capacity by adding new gNodeBs with omnidirectional or directional antennas. However, when the URLLC services require extremely low latency that do not allow for retransmissions, the system performance becomes quite sensitive to the inter-cell interference, which means that in practice the gNodeBs should be equipped with beamformed antennas to reach high levels of URLLC service capacity. Finally, an active distributed antenna system could be an efficient way to secure good coverage throughout the desired service area.

I. INTRODUCTION

When it comes to smart manufacturing, 5G New Radio (NR) defined by the Third Generation Partnership Project (3GPP) is a prime enabling technology to provide wireless connectivity in and around the factory. The global 3GPP ecosystem secures cost-efficient, future-proof products and services for evolution. Furthermore, 3GPP 5G NR can connect a variety of industrial devices with different service needs, and it can also provide ultra-reliable and low-latency communication (URLLC) to bring wireless connectivity to demanding industrial equipment [1,2]. Finally, 3GPP 5G NR supports the deployment of local non-public networks (NPN) [3] operating in licensed spectrum to ensure guaranteed quality-of-service (QoS) and tailored for the specific needs of the industrial party requesting the wireless connectivity.

When setting up a non-public factory network, one of the fundamental questions is how to design the radio network deployment so that the system can successfully serve the total offered traffic within the desired coverage area. Densification of the network deployment, for example by adding new sectors or sites within an existing communication network, has traditionally been an efficient way to improve the system coverage and capacity. It is well known that network densification can often reduce the maximum cell capacity, but as the number of cells serving a certain geographical area is increased, the overall maximum system capacity is improved. Network densification has often a negative impact on the observed geometries (i.e., downlink signal-to-interference ratio for a fully loaded network). This is due to the fact that the reduced distances between the base stations (called as gNodeBs in 3GPP 5G NR) and the users together with the increased number of neighboring cells will increase the level of the received inter-cell interference faster compared to the received signal power from the serving base station, resulting in worse signal-to-interference-plus-noise ratios (SINR) for the users, as demonstrated for example in [4-7]. Assuming a fixed level of offered area traffic (expressed as Mbps/m$^2$) instead, network densification will reduce the level of the offered traffic per cell. That can then potentially lead to a lower level of the average inter-cell interference and improved SINR values in particular if the impact of traffic offloading is large enough to compensate for the impact of the reduced geometries. Traffic offloading will also allow the users to obtain more radio resources, which has a positive impact on the user throughputs and reduces the negative impact of the worse SINRs.

In the case of URLLC with guaranteed QoS, the overall system capacity is typically determined by the worst-case performance. When it comes to smart manufacturing, a large number of URLLC use cases with different characteristics and QoS requirements have been defined. For example, the required latency bounds can vary from 1 ms to tens of ms. The different QoS requirements, with different possibilities to perform retransmissions, can then considerably affect the feasibility of the different deployment options to densify the network.

The objective of this paper is to evaluate the impact of network densification on the performance of a non-public URLLC factory network and to verify how the situation will change together with the applied QoS requirements. The rest of this paper is organized as follows. The system model and the main simulation parameters are presented in Section II. The obtained performance evaluation results are presented and analyzed in Section III. Finally, conclusions are drawn and some of the remaining open questions are discussed in Section IV.

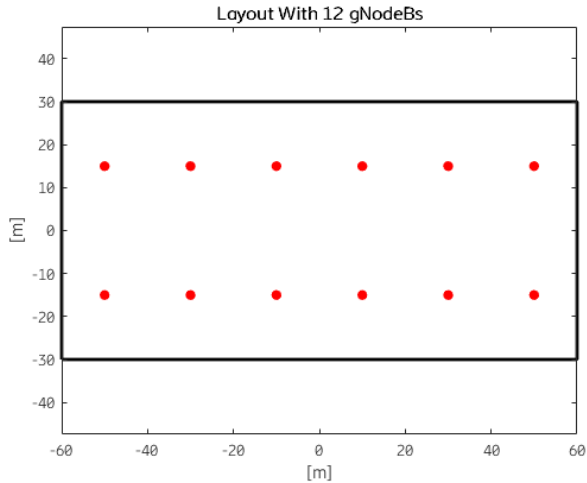

Fig. 1. Assumed network layout with 12 ceiling-mounted gNodeBs.

## II. System Model

### A. Evaluated Deployment Options

The evaluations in this paper consider a non-public URLLC network deployed within a single-floor factory building with the size of 120 m x 60 m x 10 m, see Fig. 1. In order to evaluate the impact of network densification, the URLLC network is assumed to consist of {1, 2, 3, 6, 12, 18} ceiling-mounted gNodeBs. The assumed propagation model is based on the 3GPP model for Indoor Factory with Dense clutter and High base station height (InF-DH) [8]. Furthermore, the factory floor plan and the network layout has been defined with the following parameter values adjusting the characteristics of the propagation model: clutter density ($r$) equal to 60%, effective clutter height ($h_c$) equal to 6 m, typical clutter size ($d_{clutter}$) equal to 2 m, gNodeB antenna height ($h_{BS}$) of 8 m and user equipment (UE) antenna height ($h_{UT}$) equal to 1.5 m. Finally, the performance evaluations are done for the following gNodeB antenna options:

- Omnidirectional antenna with gain equal to 2 dBi and vertical half-power beamwidth equal to 80°. No downtilt has been applied.

- Directional (sector) antenna with gain equal to 6 dBi and half-power beamwidth equal to 90° both in horizontal and vertical direction. The antenna is assumed to be pointing downwards, meaning that a 90° mechanical downtilt has been applied.

- Beamformed antenna with fully digital UE-specific beamforming (long term wideband eigen beamforming). Antenna array (4x4x2) consists of 16 cross-polarized antenna elements with gain equal to 5 dBi and half-power beamwidth equal to 90°. The antenna is assumed to be pointing downwards.

As a comparison, the performance of an active distributed antenna system (DAS) is evaluated as well. In this case, the evaluated URLLC network is assumed to consist of only one gNodeB connected to eight omnidirectional ceiling-mounted antennas with gain equal to 0 dBi, half-power beamwidth equal to 90° and an electrical downtilt equal to 35°.

### B. URLLC Performance Model

For the URLLC performance, the focus is on what latency can be provided with high reliability. This means that the achievable latency is largely determined by the worst-case timing, e.g., when the downlink packet arrives exactly at the beginning of an uplink slot or vice versa, resulting in the largest possible alignment delay. In general, when evaluating the one-way latency, the downlink transmission consists of: the gNodeB transmitting a physical downlink control channel (PDCCH) to schedule a physical downlink shared channel (PDSCH), and the UE decoding the PDSCH and sending a corresponding hybrid automatic repeat request acknowledgment (HARQ-ACK) feedback which may trigger a HARQ retransmission of the same transport block. For the configured grant (CG) uplink transmission assumed in this paper, initial transmission of the physical uplink shared channel (PUSCH) is according to the configured uplink grant without the need of sending a scheduling request, while the further PUSCH retransmission is based on dynamic scheduling from the gNodeB. Latency evaluation is done based on the time involved in the transmission process described above including all relevant processing delays in the gNodeB and UE, and the alignment delay with respect to the transmission opportunities defined by the applied time division duplex (TDD) pattern, 3GPP 5G NR release and the NR numerology. A more detailed description of the assumed latency evaluation methodology and principles can be found in [9-11].

The reliability model is constructed as a combination of both link-level simulations defining the performance (i.e., the error probability as a function of the signal-to-noise ratio (SNR)) of the different control and data channels and analytical expressions to evaluate the total reliability [12]. Based on the assumed model and the data transmission flow described above, the total downlink reliability at a given modulation and coding scheme (MCS) and SNR after $N$ transmission attempts can be expressed as given in (1). In this expression, the data transmission attempts are assumed to be correlated with each other, while the downlink control transmissions are assumed to be uncorrelated with each other and with data. This can be motivated by e.g., using different downlink control resources between transmission attempts. Furthermore, in uplink with CG, the total reliability can be calculated as given in (2). In (1) and (2), $p_1$ is the probability of successfully receiving the PDCCH, $p_{2,k}$ is the probability of a data block sent over PDSCH or PUSCH being correctly received after $k$ transmissions are soft-combined conditioned that previous $k-1$ transmission attempts fail, $p_3$ is the probability of PUCCH NACK being detected, and $p_4$ is the probability of the gNodeB detecting that a PUCCH was not transmitted [12]. The link-level simulations and the total reliability model are used to derive a model for the raw spectral efficiency as a function of the SINR, assuming the maximum allowed number of transmission attempts ($N$) within the latency bound, and using the highest MCS supporting the reliability requirement. This model will then be used as an input for the system-level URLLC capacity evaluations.

$$p_t^{DL} = \sum_{n=1}^{N} \sum_{i=1}^{n} \left\{ \binom{n-1}{n-i} [(1-p_1)p_4]^{n-i} p_1 p_{2,i} \prod_{j=1}^{i-1} p_1 p_3 (1-p_{2,j}) \right\} \quad (1)$$

$$p_t^{UL} = p_{2,1} + (1-p_{2,1}) \sum_{n=2}^{N} p_1 p_{2,n} \prod_{i=2}^{n-1} (1-p_1 p_{2,i}) \quad (2)$$

TABLE I. MAIN SYSTEM-LEVEL SIMULATION PARAMETERS

| Parameter | Value |
|---|---|
| Frequency [GHz] | 3.6 |
| Channel bandwidth [MHz] | 100 |
| Subcarrier spacing [kHz] | 30 |
| TDD pattern | DUDU, 7 OS |
| Downlink:uplink traffic ratio | 1:1 |
| QoS requirements | Latency: 1 ms, 3 ms<br>Reliability: 99.999%<br>Service availability: 100% |
| gNodeB transmit power [dBm] | 30 (gNodeB),<br>30 (active DAS, per antenna) |
| UE transmit power [dBm] | 23 |
| gNodeB antenna gain [dBi] | See Section II.A for details |
| UE antenna gain [dBi] | 0 dBi (isotropic) |
| gNodeB noise figure [dB] | 5 (gNodeB),<br>19 (active DAS) |
| UE noise figure [dB] | 9 |
| Uplink power control | SNR target = 10 dB, α = 0.8 |

Finally, the maximum URLLC capacity model is based on the performance evaluation methodology option 2 defined in [2]. The model assumes periodic URLLC traffic and that the data arrives at the same time for all the users. According to the model, the URLLC users are assumed to be successfully served if they can fulfill the desired reliability requirement within the given latency bound for both the downlink and the uplink. In practice, the desired QoS cannot be guaranteed if a) the maximum achievable user bit rate is less than what would be required to transmit the message payload during one transmission time interval (TTI), or b) the system does not have enough radio resources to successfully serve the total offered traffic load. For the URLLC performance evaluation, the following metrics have been defined: URLLC service availability (defined as the percentage of locations within the factory floor where the desired QoS can be guaranteed) and URLLC service capacity (defined as the maximum offered system traffic, at which the desired 100% URLLC service availability can still be reached).

*C. Simulation Parameters*

When it comes to the evaluated URLLC service and network, a subcarrier spacing of 30 kHz and payload packet size of 32 bytes are assumed. A TTI length of 0.25 ms is considered with 7 OFDM symbols (OS) per TTI. Furthermore, the link-level simulations have been performed for QPSK, 16 QAM and 64 QAM modulation schemes with the corresponding (1/20, 1/10, 1/5, 1/3), (1/3, 1/2, 2/3) and (2/3, 3/4) code rates, respectively.

The main system-level simulation parameters have been listed in Table I. The URLLC traffic is assumed to be periodic and symmetric between the downlink and the uplink. Furthermore, the different latency requirements can be assumed to correspond to URLLC services with different packet transmission rates. The applied TDD pattern is assumed to consist of alternating downlink and uplink sub-slots (DUDU) with 7 OFDM symbols per sub-slot. Finally, the URLLC users are assumed to be successfully served if they can fulfill the reliability requirement of 99.999% within a one-way latency bound of 1 ms or 3 ms for both the downlink and the uplink.

### III. EVALUATION RESULTS

Results from the worst-case latency evaluation are shown in Table II, assuming the URLLC features specified in 3GPP NR Release 15. As can be noticed, the assumed TDD pattern and the NR numerology can support the very stringent URLLC service with a one-way latency bound of 1 ms. However, in that case only one transmission attempt is allowed, which means that the data packet must be encoded with a very low and robust code rate, typically resulting in a low spectral efficiency. If a more relaxed URLLC service with a latency bound of 3 ms is assumed instead, a total of three transmission attempts are possible. Hence, for that kind of a URLLC service it can be more efficient to use a higher and less robust initial code rate, and to perform retransmissions based on HARQ feedback in case the initial transmission fails.

A summary of the obtained maximum URLLC service capacity results is shown in Fig. 2, assuming a one-way latency bound of 1 ms. In the figure, the maximum URLLC service capacity is expressed as the minimum of the maximum downlink and uplink service capacities. Furthermore, it can be highlighted that the maximum URLLC service capacity of the deployments with one gNodeB, including the active DAS, is uplink-limited, while the deployments with multiple gNodeBs are downlink-limited. The results demonstrate that the performance of the stringent URLLC service is very sensitive to the inter-cell interference: Network densification with omnidirectional or directional gNodeB antennas will result in a worse maximum URLLC service capacity compared to a single cell deployment with only one gNodeB located in the middle of the factory floor, or an active DAS with one gNodeB connected to eight antennas. Even though the active DAS can provide better downlink SINRs compared to the deployment with one gNodeB, the uplink SINRs and the corresponding system capacities look quite similar due to the applied power control and the fact that the propagation losses are not too high to prevent the UEs to reach their SNR targets.

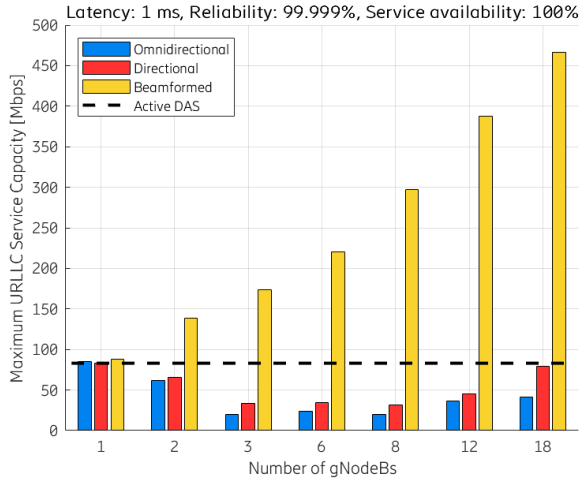 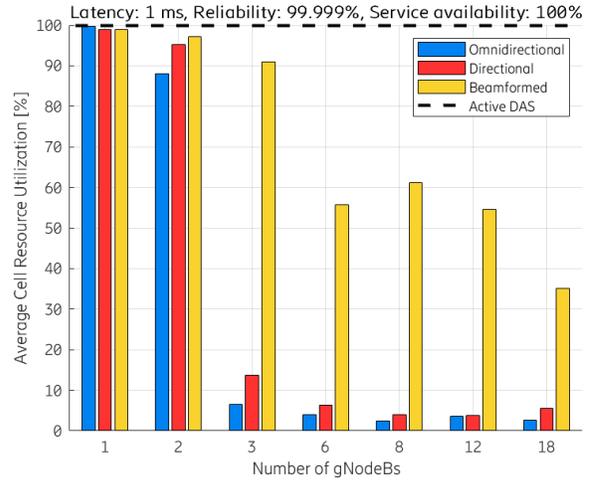

Fig. 2. Maximum URLLC service capacity for the different deployment alternatives and the average cell resource utilization for the limiting transmission direction when the offered traffic is equal to the maximum URLLC service capacity, assuming a latency bound of 1 ms.

TABLE II. DOWNLINK AND UPLINK WORST-CASE LATENCIES

| Transmission attempts | DUDU, 7 OS | |
|---|---|---|
| | *Downlink* | *Uplink* |
| Initial transmission | 0.93 ms | 0.93 ms |
| First retransmission | 1.93 ms | 1.93 ms |
| Second retransmission | 2.93 ms | 2.93 ms |

As mentioned, the assumed latency bound of 1 ms allows only one transmission attempt, requiring fairly high SINRs even for the worst users to guarantee the desired level of reliability. In other words, since a 100% service availability is required, the maximum URLLC service capacity becomes very sensitive to the occasional inter-cell interference peaks. In practice, the URLLC network should be planned to cope with the worst-case interference scenarios, which would require the use of a large interference margin when designing the network. Alternatively, the average cell load should be kept at a low level to limit the probability of the situations with a high level of the inter-cell interference. Taking a closer look at the results for the average cell resource utilization shown in Fig. 2 it becomes clear that the deployment alternatives having only one gNodeB (and hence, no inter-cell interference) can be loaded almost to 100% while still being able to guarantee the desired QoS for all users. As soon as the network has multiple gNodeBs, the level of the maximum allowed cell load (i.e., the maximum allowed average cell resource utilization) has to be limited, both to limit the probability of the high inter-cell interference peaks and to guarantee a sufficient amount of available radio resources even for the worst users. The results indicate that in the case of gNodeBs with omnidirectional or directional antennas, the maximum allowed cell load becomes very low, 15% or even considerably less, when the network has more than two gNodeBs. That will then directly result in a considerably reduced maximum URLLC service capacity, even though the total offered traffic will be shared by an increased number of cells.

The situation becomes different if the gNodeBs have beamformed antennas. For that kind of a deployment, network densification by adding new gNodeBs is clearly beneficial, and improves the overall URLLC service capacity. The main reason why the beamforming is so beneficial is that the narrow UE-specific beams are very efficient in limiting the level of the inter-cell interference, even when the cells become smaller. As a result, the geometries experienced by the users are improved and the average resource utilization required to serve a certain offered traffic is reduced much faster together with the network densification compared to the omnidirectional antennas. In all, the users will experience much better SINRs, as demonstrated by the results in Fig. 3. There, the total offered downlink traffic is fixed at 80 Mbps, which is close to the (uplink-limited) maximum URLLC service capacity for the single cell deployments. Looking at the SINR distributions for the case with one gNodeB, the beamformed gNodeB can provide slightly better downlink SINRs compared to the omnidirectional gNodeB. However, keeping in mind that the highest MCS is reached already when the SINR is around 25 dB, the deployments cannot benefit from the very high SINR values, and hence, they will have very similar performance with each other. But when the networks are densified, the performance gap between the two deployments increases rapidly.

With beamformed gNodeB antennas, most of the users experience better SINR than in the case of a single cell and their SINRs are in general improved as the network is densified, but due to the MCS limitation they do not experience any higher bit rates. Even though the worst SINRs are not as good as in the case of the single cell deployment, they are still at a much higher level compared to the deployment with omnidirectional antennas. Furthermore, even the worst SINRs (and the corresponding user bit rates) are improved as a result of network densification when the network has more than three gNodeBs. In all, network densification results in lower cell resource

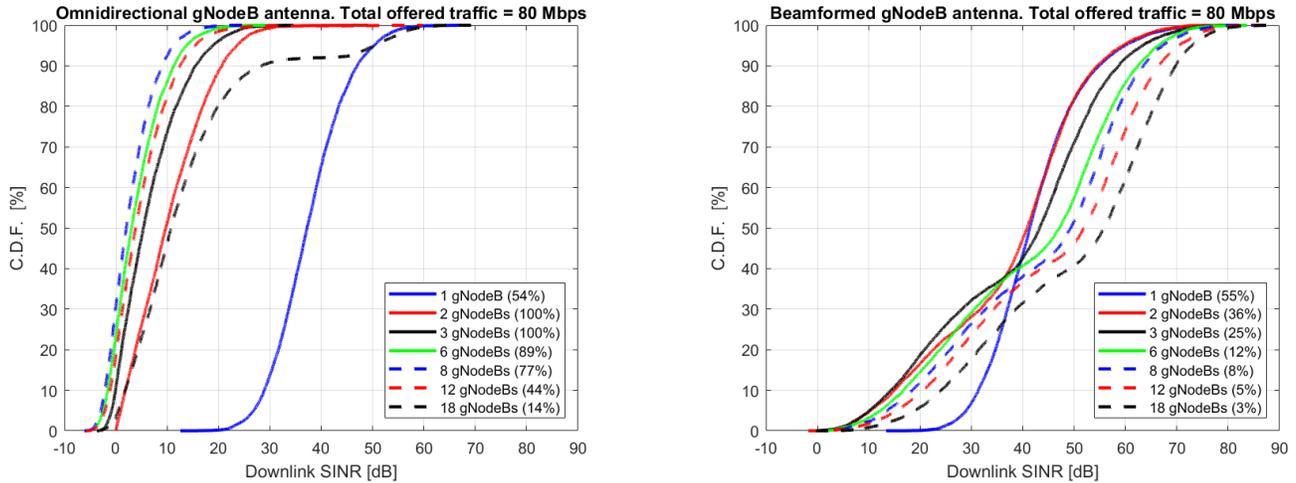

Fig. 3. Distribution of the downlink SINR values for a deployment with omnidirectional or beamformed gNodeBs, assuming a total offered traffic of 80 Mbps within the factory and a reliability requirement of 99.999% within a latency bound of 1 ms. The corresponding downlink average cell resource utilization levels are also listed.

utilization, and the URLLC network can provide the desired 100% service availability.

The situation is the opposite for the deployment with omnidirectional antennas, where the network densification results in considerably reduced SINRs. Furthermore, the reduced SINRs contribute to increased cell resource utilization and in the end, to reduced service availability (all the way down to 87% with eight gNodeBs, increasing then to 99% with 18 gNodeBs). Hence, from the SINR and service availability point of view, network densification does not have any positive impact until the network has more than eight gNodeBs.

Finally, as demonstrated by the results in Fig. 4, the maximum system capacity of the more relaxed URLLC service with a latency bound of 3 ms can be improved also by adding new gNodeBs with omnidirectional or directional antennas. However, similar to the URLLC service with a stringent latency bound, it is still considerably more efficient to densify the network with beamformed gNodeBs. With multiple transmission attempts, a maximum of three in this case, the desired level of reliability can be guaranteed with a lower SINR than in the case of a single transmission attempt, which makes the system much more tolerant to the occasional inter-cell interference peaks. As a result, higher levels of the average cell resource utilization can be allowed, leading to a higher maximum URLLC service capacity.

## IV. CONCLUSIONS AND FUTURE WORK

Densification of the network deployment, for example by adding new sectors or sites within an existing mobile communication network, has traditionally been an efficient way to improve the system coverage and capacity. That will be the case even for URLLC deployments, but the overall situation and the feasibility of the different radio network deployment options will depend on the characteristics of the URLLC service in question.

The evaluation results shown in this paper demonstrate that the URLLC services with relaxed latency requirements, allowing multiple transmission attempts, can tolerate a decent level of inter-cell interference while still being able to guarantee the desired QoS for all users. This then also makes it possible to improve the URLLC service capacity even by adding new gNodeBs with omnidirectional antennas. However, the obtained capacity gain is very modest compared to the use of directional antennas. Finally, when beamforming is applied, network densification can result in significant capacity gains.

When the URLLC services require extremely low latency that do not allow for retransmissions, the system performance becomes very sensitive to inter-cell interference. To guarantee the desired QoS for all users, actions are needed to mitigate the increasing interference caused by network densification, which can in the end be harmful from the overall system performance point of view. In the case of omnidirectional and directional antennas, the URLLC service capacity may even decrease as a result of network densification until the network becomes dense enough. Hence, in practice the gNodeBs should be equipped with beamformed antennas to reach high levels of URLLC service capacity. However, if the main reason to densify the network is to secure good coverage throughout the factory floor and not to achieve extreme system capacity, an active DAS would be a good alternative to deploying additional gNodeBs.

There are still a few open questions that should be investigated further. To start with, the impact of network densification could be evaluated for some other types of propagation environments than just the InF-DH assumed in this paper. For example, it would be interesting to perform a similar evaluation for a realistic factory environment with objects ("clutter") blocking the received signal. In such environment, it could potentially be possible to design the gNodeB locations so that the clutter is utilized to attenuate the inter-cell interference as well. Furthermore, the impact of different TDD patterns and non-symmetric downlink:uplink traffic ratios could be evaluated as well. Finally, it would be interesting to evaluate a scenario where the factory network is deployed on the mmWave band (e.g., 26 GHz) instead of the mid-band assumed in this paper. Based on the initial discussion and performance evaluation results presented in [13], it is clear that due to the challenging propagation conditions, it becomes more difficult to provide good coverage throughout the entire factory floor, meaning that a densified network deployment would be needed to secure the desired service availability, leading to similar inter-cell interference problems compared to the mid-band. Another difference compared to the mid-band is that it would perhaps be less realistic to assume large antenna arrays with fully digital UE-specific beamforming. Hence, networks operating on mmWave bands might have to rely on the use of smaller antenna arrays, analog beamforming or directional sector antennas. However, the exact impact of such radio network deployment options on the URLLC performance is still unclear.

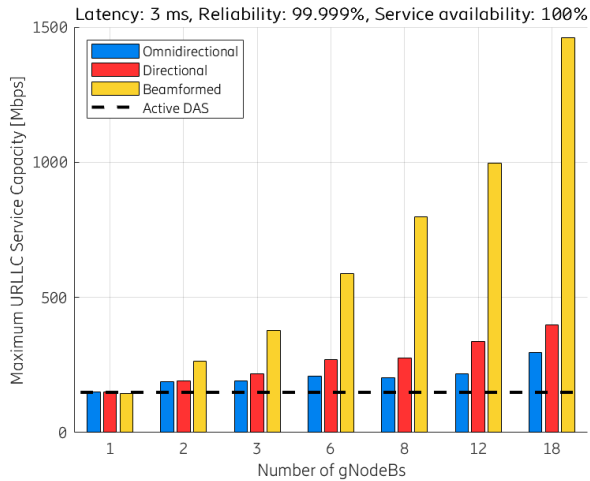 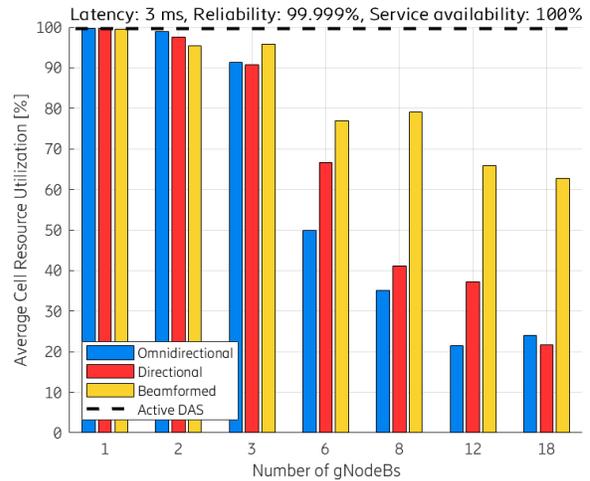

Fig. 4. Maximum URLLC service capacity for the different deployment alternatives and the average cell resource utilization for the limiting transmission direction when the offered traffic is equal to the maximum URLLC service capacity, assuming a latency bound of 3 ms.

ACKNOWLEDGMENT

This work has been performed in the framework of the H2020 project 5G-SMART co-funded by the EU. The authors would like to acknowledge the contributions of their colleagues. This information reflects the consortium's view, but the consortium is not liable for any use that may be made of any of the information contained therein.